\documentstyle[11pt,newpasp,twoside,epsf]{article}
\def\edcomment#1{\iffalse\marginpar{\raggedright\sl#1\/}\else\relax\fi}
\reversemarginpar

\begin{document}
\title{The cosmological evolution of BL Lacs: The REX point of view}
 \author{A. Caccianiga}
\affil{Observat\'orio Astron\'omico de Lisboa, Lisbon, Portugal}
\author{T. Maccacaro, A. Wolter, R. Della Ceca}
\affil{Osservatorio Astronomico di Brera, Milan, Italy}
\author{I.M. Gioia}
\affil{Istituto di Radioastronomia del CNR, Bologna, Italy}

\begin{abstract}
We present the results on the cosmological evolution of BL Lac objects
as derived from a statistically complete sample of 44 BL Lacs selected 
from the X-ray bright tail of the REX survey. 
With this sample, we have investigated the cosmological properties of
BL Lacs taking into account the radio, optical 
and X-ray limits. We infer that no evolution is clearly visible down
to the flux limits reached by our sample. On the other hand, 
deeper samples are probably needed in order to
detect the negative evolution found in the EMSS sample. The identification
of such deeper sample, extracted from the REX survey, is in progress. 

\end{abstract}

\section{Introduction}

The first study of the evolutionary behavior of BL Lac objects
was based on EMSS data. By studying the surface density 
of the BL Lacs as a function of the flux (Maccacaro et al. 1989, 
Wolter et al. 1991) and by applying the $V_e/V_a$ test 
(Morris et al. 1991; Wolter et al. 1994), evidence for negative 
evolution (objects
less numerous/luminous in the past) has been found. At the same time,
the analysis of the 1 Jy radio selected sample of BL Lacs did 
not show any strong evidence
for cosmological evolution (Stickel et al. 1991) in contradiction with
what had been found in the X-ray band. These results
were based on small samples containing 22 and 34 objects respectively.
Because of the difficulty of selecting larger complete samples of BL Lacs
a revision of their cosmological evolution is still a difficult task. 
Nevertheless, the recent availability of large radio and X-ray surveys boosted
the attempts of selecting large sample of BL Lacs (e.g. 
the REX survey, Maccacaro et al. 1998, Caccianiga et al. 1999, the DXRBS, 
Perlman et al. 1998, the RGB, Laurent-Muehleisen et al. 1998). 
Using the ROSAT All Sky Survey, Bade et al. (1998) selected a complete sample
of 33 BL Lacs with a flux limit of 8$\times$10$^{-13}$ ergs cm$^{-2}$ s$^{-1}$
in the 0.5-2.0 keV band. They have applied the $V/V_{max}$ test to compute the 
cosmological evolution of BL Lacs, finding hints of negative evolution
($<V/V_{max}>$=0.40 $\pm$ 0.06). Moreover, they have found that the negative
evolution depends on the optical-to-X-ray spectral index ($\alpha_{OX}$)
being more extreme ($<V/V_{max}>$=0.34 $\pm$ 0.06) for ``X-ray'' extreme 
($\alpha_{OX}\leq$0.91) objects. A negative evolution has been found 
also by Giommi, Menna \& Padovani (1999) in a sample of HBLs 
($\alpha_{RX}\leq$0.56).  
The analysis of this sample gives evidence for a negative evolution 
which depends on the radio flux limit, being stronger for radio flux
limit of 3.5 mJy ($<V_e/V_a>$=0.42$\pm$0.02) and consistent with
no-evolution for a flux limit of 20 mJy (($<V_e/V_a>$=0.49$\pm$0.04).
In this paper we present an independent study of the cosmological
evolution of BL Lacs based on the REX survey.

\section{The X-ray bright REXs}
The REX survey is the 
result of a positional cross-correlation between the  
NRAO VLA Sky Survey (NVSS, Condon et al. 1998) at 1.4 GHz and
an X-ray catalogue of about 17,000 serendipitous sources detected
in $\sim$1200 pointed ROSAT PSPC fields.
The flux limit in the radio band (at 1.4 GHz) is  5 mJy. 
In the X-ray band the flux limits range from $\sim$3.5$\times$10$^{-14}$ 
erg s$^{-1}$ cm$^{-2}$ to $\sim$2$\times$10$^{-13}$ erg s$^{-1}$
cm$^{-2}$ in the 0.5--2.0 keV band. The area covered at 
the highest flux limit is about 2200 deg$^2$.
The cross-correlation has produced a catalogue of $\sim$1600 Radio Emitting
X-ray sources (REXs).
The spectroscopical identification of the sample is in progress and, 
so far, about 36\% of the sample has been identified.
In this paper, we present a complete sample, 
the X-ray Bright REX sample (XB-REX), selected from the REX survey 
with the following additional criteria:

1) X-ray flux ($f_X$) in the 0.5-2.0 keV band $\geq$ 4$\times$10$^{-13}$
erg s$^{-1}$ cm$^{-2}$;

2) Magnitude (APM O) brighter than 20.4 

The radio flux limit is the same of the whole REX survey (5 mJy). 
The resulting sample contains 190 objects. About 93\% of these objects have
been already identified, either from literature or from our own 
spectroscopy. 
The classification of a BL Lac is based on the following criteria:
1) Equivalent width of any emission lines $\leq$ 5\AA; 
2) Ca II contrast at 4000\AA\ (B) $\leq$ 40\%.
The last  criterium is an extension of the ``classical'' one (B$\leq$25\%),
firstly used by Stocke et al. (1989). There are many evidences, in fact,
that the limit of 25\% is too restrictive and can miss some true
BL Lac objects. 

In total, the XB-REX sample contains 44 BL Lacs. Given its relatively high
X-ray flux limit  we do not 
expect to significantly detect the cosmological evolution found in 
the EMSS sample. In fact, the X-ray flux limit of the XB-REX sample
corresponds to a flux limit of  7.2$\times$10$^{-13}$ erg s$^{-1}$ 
cm$^{-2}$ in the 0.3-3.5 keV band  (the Einstein IPC energy band) 
assuming $\alpha_X$=1, which is higher than the deepest flux limit
of the EMSS complete sample of BL Lacs (f$_{(0.3-3.5)}=$5$\times$10$^{-13}$
erg s$^{-1}$ cm$^{-2}$). The flux limit of the XB-REX is in the 
region where the LogN-LogS produced from the EMSS survey 
(Wolter et al. 1991) starts to flatten and, thus, we are probably not
sampling the region where the negative evolution is more evident.

\section{The $V_e/V_a$ analysis}

In order to compute the cosmological evolution of BL Lacs
we have applied the $<V_e/V_a>$ method described in Avni \& Bachall (1980)
where the $V_a$ is the smallest ``avaliable'' volume  
among the $V_a$ computed in the three selection bands (X-ray, optical, radio).
For the objects without a redshift, 
we have assumed z=0.27 (the mean value for the sample). We have also assumed 
$\alpha_R$ = 0, $\alpha_O$=1.5 and $\alpha_X$ = 1 for the spectral indices 
in the radio,  optical and X-ray bands, respectively.
The resulting value of the $<V_e/V_a>$ is reported in Tab.~1.
\begin{table}
\begin{center}
\caption{Results of the $V_e/V_a$ analysis}
\label{results}
\begin{tabular}{l l l l r }
\hline
Sample & N. of objects & $<V_e/V_a>$ value & Error$^{a}$ & KS prob.(\%) \\
\hline 
Total                 & 44 & 0.52 & 0.04 & 94 \\
``classical'' BL      & 35 & 0.50 & 0.05 & 84 \\      
$\alpha_{OX}\leq$0.91 & 16 & 0.52 & 0.07 & 91\\
$\alpha_{RX}\leq$0.62 & 20 & 0.53 & 0.06 & 97 \\
EL AGNs               & 78 & 0.64 & 0.03 &  2\\

\hline
\end{tabular}
$^{a}$1/$\sqrt{12N}$, where N=number of objects
\end{center}
\end{table}

The X-ray band is the limiting one in many 
cases (20) but also the radio and the optical bands play an 
important role (they are the limiting band in 13 and 11  cases,
respectively).
The K-S test shows that the $<V_e/V_a>$ values are uniformly distributed 
between 0 and 1 (probability=94\%). 
The result, as expected, is not significantly affected by the missing z. 
If we ignore the objects without z we obtain $<V_e/V_a>$ = 0.52 $\pm$ 0.05
while if we assign to all the objects without a measured z the maximum 
value observed in 
the sample (z=0.6) we obtain  $<V_e/V_a>$ = 0.54 $\pm$ 0.04.

We have then divided the sample in different ways and applied 
the $V_e/V_a$ analysis to these sub-samples. The results are
reported in Table~1. 
First of all, we have investigated whether the ``extended'' criteria used to 
define a  BL Lac in the XB-REX sample, namely the Ca break below 40\%
instead of 25\%, affects the result of the $<V_e/V_a>$ analysis. To this
end, we have computed the $<V_e/V_a>$ by using only the ``classical''
BL Lacs, i.e. defined with the most restrictive limit on the 
Ca break (25\%). The 35 ``classical'' BL Lacs still have a $<V_e/V_a>$
consistent with no-evolution. 
We have then checked if there is any dependence of the $<V_e/V_a>$ value
on the ``type'' of BL Lac. We have thus considered  only the 17 BL Lacs 
with an ``extreme'' X-ray/optical ratio ($\alpha_{OX}\leq$0.91).
According to the results
found by Bade et al. (1998) these sources should show an extremely 
negative evolution. Instead, we do not see any evidence for negative
evolution and the $V_e/V_a$ value is still consistent with 0.5.
We have also divided the sample according to the ratio between the
X-ray and the radio flux and considered only the typical 
X-ray selected BL Lacs (HBL type) with $\alpha_{RX}\leq0.62$. 
Again, the resulting $<V_e/V_a>$ is consistent with 0.5. 

For comparison, we have computed the $V_e/V_a$ for the 78 Emission Line
AGNs found in the XB-REX sample. 
In this case, the most limiting band is the X-ray one (60 objects limited
by the X-ray) while the optical and the radio bands play a marginal role. 
The result is fully consistent with what found in the EMSS sample 
($<V_e/V_a>$=0.62 $\pm$ 0.01, Della Ceca et al. 1992). The distribution of the
$V_e/V_a$ values  is clearly not uniform (KS probability of 2\%).

\section{Discussion and conclusions}

The analysis of the complete sample of 44 BL Lacs presented here shows no 
evidence for cosmological evolution down to the explored flux limits. 
Even if we restrict the
analysis to the most ``X-ray extreme'' objects (i.e. $\alpha_{OX}\leq$ 0.91
or $\alpha_{RX}\leq$0.62), we do not find any sign of negative 
evolution. 
As stated before, the XB-REX sample is probably not deep enough to 
see the negative evolution detected in the EMSS survey.
Instead, the RASS sample has an X-ray limit a 
factor 2 higher than the XB-REX sample and the negative evolution 
found by Bade et al. (1998) should have been detected in our analysis.
In any case, a deeper sample will be instrumental to assess if a 
negative evolution affects the population of BL Lacs. At a flux limit 
2.8$\times$10$^{-13}$ erg s$^{-1}$ cm$^{-2}$  (0.5-2.0 keV band), 
which corresponds to the lowest limit in the 
EMSS-C sample, the REX survey is identified at the 80\% level, 
with the same 
radio and optical constraints of the XB-REX sample presented here. After the 
identification of the remaining sources is completed, 
we should be able to make a firmer
statement on the cosmological evolution of the sample. 

{\bf Acknowledgments} This research was partially supported by the European 
Commision, TMR Programme XCT96-0034 ``CERES'', by the FCT under grant
PRO15132/1999, by Italian Space Agency (ASI) and the Italian MURST under grant
COFIN98-02-32.

\end{document}